\documentclass[preprint,12pt,3p]{elsarticle}

\usepackage{url}
\usepackage[T1]{fontenc}
\usepackage[utf8]{inputenc}
\usepackage{listings}
\usepackage{indentfirst}
\usepackage{subcaption}
\usepackage{lipsum}
\usepackage{multicol}
\usepackage{hyperref}
\usepackage{color, colortbl, soul}
\soulregister\cite7
\soulregister\ref7
\soulregister\pageref7
\soulregister\textit7
\definecolor{lightyellow}{rgb}{1,1,1}
\definecolor{lightGray}{gray}{0.9}
\sethlcolor{lightyellow}
\definecolor{yellow}{rgb}{1,1,1}
\usepackage{balance}
\usepackage{amssymb}
\usepackage{latexsym}
\usepackage{graphicx}
\usepackage{comment}
\usepackage{booktabs}
\usepackage{multirow}
\usepackage{framed}
\usepackage{comment}
\usepackage{longtable}
 \usepackage{float}
\usepackage{makecell}
\usepackage{soul}
\definecolor{Gray}{gray}{0.75}
\definecolor{LGray}{gray}{0.95}
\usepackage[obeyFinal]{todonotes}
\usepackage{comment}
\usepackage{mathtools}
\usepackage{xcolor}
\usepackage{verbatim}

\newcommand\mape[1]{{#1}}

\journal{Information Processing and Management}

\begin{document}

\begin{frontmatter}

\title{On the efficacy of old features for the detection of new bots}

\author[label1,label4]{Rocco~De~Nicola}
\address[label1]{IMT Scuola Alti Studi Lucca, Piazza San Francesco 19, 55100 Lucca, Italy}
\address[label4]{CINI Cybersecurity Lab, Via Ariosto, 25, 00185 Roma, Italy}
\ead{rocco.denicola@imtlucca.it}

\author[label2,label1]{Marinella~Petrocchi\corref{cor1}}
\address[label2]{Istituto di Informatica e Telematica, CNR, via G. Moruzzi 1, 56124, Pisa, Italy}
\ead{marinella.petrocchi@iit.cnr.it}

\author[label1]{Manuel Pratelli}
\cortext[cor1]{Corresponding author}
\ead{manuel.pratelli@imtlucca.it}

\begin{abstract} For more than a decade now, academicians and online platform administrators have been studying solutions to the problem of bot detection. Bots are computer algorithms whose use is far from being benign: malicious bots are purposely created to distribute spam, sponsor public characters and, ultimately, induce a bias within the public opinion. To fight the bot invasion on our online ecosystem, several approaches have been implemented, mostly based on (supervised and unsupervised) classifiers, which adopt the most varied account features, from the simplest to the most expensive ones to be extracted from the raw data obtainable through the Twitter public APIs. In this exploratory study, using Twitter as a benchmark, we compare the performances of four state-of-art feature sets in detecting novel bots: one of the output scores  of the popular bot detector Botometer, which considers more than 1,000 features of an account to take a decision; two feature sets based  on the account profile and timeline; and the information about the Twitter client from which the user tweets. The results of our analysis, conducted on six recently released datasets of Twitter accounts, hint at the possible use of general-purpose classifiers and cheap-to-compute account features for the detection of evolved bots.
\end{abstract}



\end{frontmatter}

\section{Introduction}
In a pioneering study on social media manipulation it is shown that, on the eve of the 2010 Massachusetts election, a handful of automated Twitter accounts spread misinformation about the Democratic candidate~\cite{mustarafi10}. Although Twitter quickly banned those accounts, the damaging information had already circulated beyond the platform. The bots had managed to create a viral cascade, and even Google search results were reporting the false news. 
Despite advancements in techniques for bot detection and search-engine filtering and ranking, automated accounts still represent a menace for societal and political debates. Bots have played an important role in every significant event: from elections (see, e.g., the 2016 and 2020 US elections~\cite{sharma2020identifying} and the Brexit referendum) to the spreading of propaganda and misinformation on hot topics like immigration and pandemics~\cite{caldarelli2020immi,caldarelli2020covid}. 
Unfortunately, in the last ten years, skills of bots have improved dramatically~\cite{Ferrara2016rise, cresci2017paradigm}, in that fake accounts are able to create a network of friends, chat with genuine accounts without betraying their nature, and act in a coordinated fashion. Their actions are no longer individual: synchronisation and coordination  characterize teams of bots that, if considered  individually, can be hardly detectable by state-of-the-art classifiers.
Quoting a pioneer in the field, `There is no shortage of research challenges, even 10 years later, to try to identify this kind of manipulation’~\cite{menzer2020webofscience}.

\mape{{\it Motivations:} 
Bots have evolved over the years: from rather simplistic accounts to sophisticated ones, whose behaviour and online interactions are almost indistinguishable from those of genuine accounts~\cite{Ferrara2016rise,cresci2017paradigm}. This paper grew out of a curiosity the authors had when reading a 2019 paper about online manipulation during the 2016 US presidential elections~\cite{bovet2019Nature}. Specifically, in order to estimate how many bots had participated in the online political discussion on Twitter, the article proposed to check the Twitter client from which tweets were posted (intuitively, unofficial clients are used by professionals to automate some tasks). Intrigued by this simple approach, we wondered  to what extent the features and methods used years ago to detect early bots are still effective, and if so, to what extent they can be used for detecting new bots. 
}

{\it Research objective:} The challenge is therefore to develop general-purpose bot detection systems, that, even if not targeted at detecting specific disinformation and spam promotions, are still able to offer good performances in detecting multiple types of bot accounts, including those belonging to strategic campaigns~\cite{DBLP:conf/aaai/YangVHM20}. 
The main objective of our research is thus to conduct an exploratory study on the efficacy -in terms of performances- of general-purpose bot detectors, also for bots that operate in a coordinated manner. We will use Twitter as a benchmark.

{\it Methodology:} To achieve this objective, we consider well known learning algorithms to see how different feature sets are effective when applied to recently
released
 Twitter accounts datasets, in which it is known a priori the nature of the accounts (either bots or genuine ones).   As a first feature, we consider one of the score provided by Botometer~\cite{Varol2017, Yang2019arming}, the popular bot detector developed at Indiana University. This tool examines more than 1,000 features of a Twitter account (concerning, for example, properties related to the network, the timeline, and the profile of the account), and returns, as output, a set of scores indicating the level of `boticity' of the account. 
The second feature set is retrievable from the profile of the account. The third feature set instead relies only on features related to the timeline of the account. Both sets disregard the features that are known to be the most expensive to compute (mainly in terms of time needed for data gathering), namely those concerning the account's relationships (friends and followers).
The fourth feature is the Twitter client used for tweeting. It has been leveraged by  Bovet \& Makse in~\cite{bovet2019Nature} to analyse a dataset of tweets concerned with the US 2016 presidential elections and filter tweets posted by bots. 

{\it Contributions:} The main contribution of this manuscript is 
an investigation of the performances of well-established learning algorithms, trained with different sets of features, on six recently released datasets of bots and human-operated accounts  
whose composition is known a priori. The main findings of this analysis are as follows:
\begin{itemize}
    \item For some of the training sets, the inspection of the Twitter client is sufficient to discriminate between bot accounts and genuine ones. However, we have to say that we have been considering bots that are very easy to recognise: they are self-declared bots whose nature their programmers have no intention of hiding;
    \item  Considering the features based on the profile and timeline of the account, they work well both for bots acting individually and for bots acting in teams. This is quite surprising. It is not the first time that so-called `low cost' features have been proven to work well for bot detection (see, e.g., the work by Cresci et al. on fake followers detection~\cite{cresci2015fame}, that focused on bots acting only in isolation). 
    \item \mape{We test 
    very recent, novel bots. The good performance obtained in the classification phase suggests that it is possible to counteract the evolution of bots by leveraging well-known features already tested on less evolved versions.}
    \item \mape{Overall, the evaluation results lead us to argue that a general-purpose bot detector, not specifically created to recognise particular types of accounts only, can be used to skim anomalous accounts, and, subsequently, specific techniques can be applied to uncover coordinated behaviour among these accounts.}
\end{itemize} 


{\it Roadmap:}
The remainder of the paper is organized as follows. Section~\ref{sec:LiteratureReview}
discusses recent related work, positioning our study among relevant state-of-art papers.
In Section~\ref{sec:datasets}, we introduce the datasets we use for the analyses.  
Section~\ref{sec:meth} provides the description of the feature sets and the experimental setup. Section~\ref{sec:exp} shows the experimental results. In Section~\ref{sec:discussion}, we discuss the main findings and the limitations of our investigation. 
Finally, Section~\ref{sec:concl} draws the conclusions. 


\section{Literature Review}
\label{sec:LiteratureReview}

This literature review first highlights the different trends that have characterized bot detection over the years. Then, it considers recent works and compares them with ours. 

We would like to highlight the trends by comparing them to the historical cycles coined by the Neapolitan philosopher Giambattista Vico, who lived at the turn of the 17th and 18th centuries.
He developed a unique theory of human history. He was convinced that history was characterised by the continuous and incessant repetition of 3 distinct cycles~\cite{vicoscienzanuova}: 
1) the \textit{age of the gods}, in which human beings believe they live under divine rule; 2) the \textit{age of the heroes}, in which they form aristocratic republics; 3) the \textit{age of mankind}, when all  finally recognise themselves as equal, because they all belong to mankind.
When we talk about bot detection, we are talking about research carried out in the last 10-11 years (indeed, the first work appeared around 2010~\cite{YardiRSB10,mustarafi10}). We are certainly not talking about centuries, as in the case of Vico's cycles. But, it appears surprising how the theory seems to apply to the field of bot detection, at least with respect to the existence of three distinct phases. 
In fact, in the first years of bots hunting (~2010-2013/14), researchers mainly focused on supervised machine learning and on the analysis of the single account: `classifiers were separately applied to each account of the group', to which they assigned a bot or not label~\cite{cresci2020decade}. This first period can be seen as the Vico's age of the gods, where the `deus ex machina' comes and solves the brand new research question.

Approximately in the last five years, say up to 2019, instead, a number of research teams independently 
have proposed new approaches
that aim at detecting coordinated and synchronized behavior
of groups of automated malicious accounts, see, e.g.,~\cite{viswanath2015, IntSys2015, yu2015}. Thus, researchers no longer took the individual account and classified it as bot or not. Instead, they considered a group of accounts, and their common characteristics, i.e., those that may show that some of them have been programmed for the same purpose (and therefore, for example, they behave online in the same way). One of the proposed techniques consists in measuring the similarity of the traces left by the accounts in their online activities~\cite{cresci2018social}: by composing the account's Digital DNA (i.e., a string of characters that encodes the actions of the account), it is possible to leverage string mining tools and bring out from the crowd accounts that behave in the same way. This second period can be seen as the Vico's age of the heroes, who have travelled new paths to the detection of teams of bots.

Surprisingly, 
after 2019, we have witnessed the confluence of the two research strands: obtaining so-called \emph{general-purpose} classifiers, i.e., classifiers that are not specific to the type of bots, but which are also able to spot bots belonging to teams which before could be detected only as part of a team and not as individuals. 
This third period can be seen as the Vico's age of mankind, where researchers join their various efforts.

In the following, we shall describe recent contributions to general-purpose bot detectors, while highlighting differences and similarities with ours. 

The recent work by Sayyadiharikandeh et al.~\cite{DBLP:conf/cikm/Sayyadiharikandeh20} starts from the observation that different bot classes have different sets of informative features and,  
thus, the authors build specialized supervised models for distinct bot classes. The specialized models are aggregated into an ensemble and their outputs are combined through a voting scheme. 
Our work does not propose a new approach for bot detection, as the one in~\cite{DBLP:conf/cikm/Sayyadiharikandeh20}. On the contrary, we adopt general techniques and features already used in the past literature. However, we see similarities because both tend to test if a classifier can identify more than one kind of bots.
On the tested datasets, the performance results are comparable and thus encouraging. 

With the goal of testing the generality of traditional detectors on different types of bots in mind, Yang et al. ~\cite{DBLP:conf/aaai/YangVHM20} propose a highly scalable framework that enables them to handle the full stream of public tweets on Twitter in real time. Key recipe of the analysis is to use account features that have a very low cost in terms of the data needed to compute them. In particular, the authors use features related only to the account profile. \mape{This type of features was used by the second author of this manuscript to detect fake Twitter  followers~\cite{cresci2015fame} in 2015, i.e., 5 years before the publication in~\cite{DBLP:conf/aaai/YangVHM20}. 
This makes us confident that the adoption of `old' features for bot detection are worthy of study, for the same purpose, nowadays.}
It would be interesting to verify the sustainability of the approach of~\cite{DBLP:conf/aaai/YangVHM20} to perform real time analysis when considering 
account features related to the timeline. In this manuscript, we consider, e.g., mentions, hashtags, and URLs in the account tweets.


Schuchard and Crooks~\cite{10.1371/journal.pone.0244309} consider a large corpus of Twitter posts related to the 2018 U.S. Midterm Election. They do not consider different sets of features on the same training set but
run three state-of-art models, namely Botometer in the version presented in~\cite{Varol2017}, DeBot~\cite{7837909} and Bot-hunter~\cite{Bothunter2018} to test their agreement. \mape{Among the accounts responsible for a volume of more than 40 million tweets, about 254k accounts were identified as bots (bots are identified as such if at least one detector classifies it as a bot). The concordance of classifiers was poor, in fact, among the aforementioned 254k accounts, rather incredibly only 8 were labelled as bots by all 3 detection methods. The latter also differed in classifying bots according to how much they interacted with human accounts. This led the authors of~\cite{10.1371/journal.pone.0244309} to recommend a mix of detection methods to apply when hunting bots in the wild.
}

Ferrara~\cite{DBLP:journals/firstmonday/Ferrara17} explores the most significant features considered in the literature to detect bot accounts and thus rely on the user profile, on the geolocalization of the account and its activity (like the number and frequency of tweets, and the ratio between tweets and retweets),  to create a simple, yet effective, bot detection system by using a variety of algorithms available in Scikit-learn~\cite{pedregosa11}.
We see similarities with~\cite{DBLP:journals/firstmonday/Ferrara17} and our work, because they both consider, for the classification process, features that are relatively easy to compute, in terms of time necessary for data gathering. Thus,
it would be interesting to evaluate the features in~\cite{DBLP:journals/firstmonday/Ferrara17} on our datasets, which are much more recent. 

El-Mawass et al.~\cite{ELMAWASS2020102317} propose a mixed approach to reduce the false positives rate in traditional supervised systems, thus increasing recall without deteriorating precision. They rely on a classifiers cascade, and use the output of supervised classifiers as prior beliefs in a probabilistic graphical model framework. This framework allows beliefs to be propagated to similar social accounts. Classification results show a significant increase in recall and a maintained precision.

The work described so far is based on the classification of the individual account (the classifier takes as input the account, and says if it is a bot or not). In the following, we would like to briefly mention work that belongs, for the most part, to the second period of bot detection, in which a group of accounts is analysed for its `team' characteristics (for example, by comparing the creation time of the accounts, or by comparing their online behaviour).
Hui et al.~\cite{DBLP:conf/icwsm/HuiYTM20} propose BotSlayer, a tool based on an anomaly detection algorithm that let coordinated campaigns emerge from the crowd by highlighting hashtags, links,  phrases, and trending media. 
 Other techniques are based, for example, on anomalies in synchronicity and normality~\cite{Giatsoglou2015,jiang2016}, detection of loosely synchronized actions~\cite{Cao:2014}, distance between distributions of reputation scores~\cite{viswanath2015}, and similarity between sequences of actions~\cite{cresci2018social}. 
 
%
Finally, Sharma et al, in~\cite{sharma2020identifying}, move away from the research trend  of detecting teams of bots on the basis of features concerning coordination and synchronous behavior between such accounts and propose an approach to automatically uncover coordinated group behaviours from account activities and interactions between accounts, based on temporal point processes. 




\section{Datasets}\label{sec:datasets}
%

We consider six publicly available datasets\footnote{Bot Repository Datasets: \url{https://botometer.iuni.iu.edu/bot-repository/datasets.html}} that have been chosen 
 on the basis of the date of the accounts' creation, namely we chose those 
 dated 2018 or 2019. This was done  to understand whether the approaches used in our comparative analysis allow for the detection of bots more sophisticated than those created years ago. The datasets are the following:

\begin{description}
    \item[\texttt{Verified}]  first introduced
    in~\cite{DBLP:conf/aaai/YangVHM20}, consists of 2,000 verified human accounts.   Twitter offers the possibility (upon request of the account owner) to obtain an official certification of account's authenticity. 
    Certified accounts are tagged as 
verified and on the official portal have a blue circle, with a white tick at the center.
    \item[\texttt{Celebrity}] first introduced in ~\cite{FerraraArming2019}, 
    consists of 5,970 celebrity accounts collected as authentic users.
    \item[\texttt{Botwiki}] 
    consists of 704 publicly declared bots from \url{https://botwiki.org}. \mape{The Botwiki organization has the goal to preserve `examples of interesting and creative online bots, providing tutorials and other resources to folks interested in making them'. The \texttt{Botwiki} Twitter accounts are easily recognisable bots, by, e.g., the word `bot’ in the username or in the account description. These bots do not hide their automatic nature or their malicious intentions. } 
    
    \item[\texttt{Rtbust}] 
    built by Mazza et al.~\cite{crescirtbust2019}  by processing all the (9,989,819) Italian retweets shared between 17 and 30 June, 2018, by 1,446,250 distinct users. After reducing the number of considered users, using the number of their retweets ratio, 
    759 accounts were labelled either as human-operated or  bot-operated.
    \item[\texttt{Stock}] first introduced by Cresci et al.~\cite{10.1145/3313184}
    consists 
    of both genuine and automated accounts tweeting so-called cashtags, i.e., specific Twitter hashtags that refer to listed companies. Part of the automated accounts have been found to act in a coordinated fashion,  
in particular by mass retweeting cashtags of low capitalization companies.
\item[\texttt{Vendor}] introduced in ~\cite{FerraraArming2019}, consists of 1,088 fake followers, namely fake account employed to inflate the number of followers of a target account. 
\end{description}

Table~\ref{tab:orig-datasets} reports the name of the datasets, their brief description, and the number of accounts they originally feature; the year represents the average of the creation years of the accounts that belong to the dataset. 
It is worth noting that the original datasets only include the IDs of the accounts, plus the tag defining their nature. Therefore, through the Twitter public APIs and the Tweepy library (\url{http://www.tweepy.org/}), we retrieved the data we needed for the experiments shown in this manuscript.
Thus, Table~\ref{tab:actual-datasets} reports statistics about the same datasets, at the time of the current study (July, 2020).  
The table contains the number of bots and humans that have at least 0, 100, 200, 300, and 400 tweets in their timelines -we always consider the most recent 100, 200, 300, and 400 tweets posted by the accounts. We can notice that the number of accounts per dataset changes by considering a minimum number of tweets in the timeline of each account. In particular, by considering only accounts with at least {\it x} tweets in their timelines, the number of accounts per dataset decreases when {\it x} increases. This aspect is important to evaluate the performance of a classifier.
In the Discussion section, to assess the robustness of our results, we will compare how the classification performance is affected by the number of tweets (and, consequently, the number of accounts) we consider. 

\mape{For the sake of completeness, we highlight also that, if we focus on considering at least 500 tweets, instead of 400,  we lose between 2\% and 5\% of accounts, depending on the dataset. The loss is not drastic and, on the one hand,  we could have analysed accounts with a larger number of tweets in their timeline.  On the other hand, we prefer to fix the minimum number of tweets to consider equal to 400, to maintain a good trade-off between number of tweets and number of accounts}.


\begin{table*}[ht]
	\scriptsize
	\centering
	\begin{tabular}{llrrc}
 		\toprule
		\textbf{dataset} & \textbf{description} & \#bot & \#human & year \\
		\midrule
		\texttt{Verified} & verified Twitter accounts & 0 & 2000 & 2019 \\
		\texttt{Celebrity} & celebrities & 0 & 5970 & 2019 \\
		\texttt{Botwiki} & self-declared bots & 704 & 0 & 2019 \\
		\texttt{Rtbust} & genuine accounts/coordinated bots  & 391 & 368 & 2019 \\
		\texttt{Stock} & genuine accounts/coordinated bots  & 18508 & 7479 & 2018 \\
		\texttt{Vendor} & fake followers & 1088 & 0 & 2019 \\
		
		\bottomrule
	\end{tabular}
	\caption{\small Description of the original datasets.
	\label{tab:orig-datasets}}
\end{table*}

\begin{table*}[ht]
	\scriptsize
	\centering
	\begin{tabular}{lrrrrrrrrrrrrrr}
		\toprule
		& \multicolumn{2}{c}{{\textbf{\#post$\geqslant$0}}}&& \multicolumn{2}{c}{{\textbf{\#post$\geqslant$100}}}&& \multicolumn{2}{c}{{\textbf{\#post$\geqslant$200}}}&& \multicolumn{2}{c}{{\textbf{\#post$\geqslant$300}}}&& \multicolumn{2}{c}{{\textbf{\#post$\geqslant$400}}}\\
		\cmidrule{2-3}
		\cmidrule{5-6}
		\cmidrule{8-9}
		\cmidrule{11-12}
		\cmidrule{14-15}
		\textbf{dataset} & bots & human && bots & human && bots & human && bots & human && bots & human \\
		\midrule
		\texttt{Verified} & 0 & 1974 && 0 & 1965 && 0 & 1952 && 0 & 1928 && 0 & 1919\\
		\texttt{Celebrity} & 0 & 5649 && 0 & 5180 && 0 & 4902 && 0 & 4621 &&  0 & 4375\\
        \texttt{Botwiki} & 664 & 0 && 654 & 0 && 649 & 0 && 646 & 0 &&  642 & 0 \\
        \texttt{Rtbust} & 315 & 317 && 315 & 315 && 315 & 314 && 309 & 310 &&  303 & 308\\
        \texttt{Stock} & 6842 & 5880 && 6594 & 5021 && 6201 & 4390 && 5914 & 4043 &&  5619 & 3888\\
        \texttt{Vendor} & 699 & 0 && 298 & 0 && 243 & 0 && 229 & 0 &&  217 & 0\\
        
		\bottomrule
	\end{tabular}
	\caption{\small Number of accounts with a number of posts $\geqslant$ a given threshold. Timelines were crawled on July 2020 through the Tweepy library.
		\label{tab:actual-datasets}}
\end{table*}


\section{Feature sets}\label{sec:meth}
In this section, we consider four different feature sets that can be given as input to learning models for the detection of bot accounts. 
First, we  consider as feature one of the scores output of Botometer~\cite{Varol2017,FerraraArming2019}, the popular
bot detection tool developed at Indiana University.  Then, we describe two feature sets, firstly introduced by Cresci et al. in~\cite{cresci2015fame},  that require low computational resources for data gathering. Finally, inspired by Bovet and Makse~\cite{bovet2019Nature}, we take into account the type of Twitter client (official/unofficial) from which the account tweets the most.
%

\subsection{CAP\_UNI Botometer score}\label{sec:botometer}
Botometer v3~\cite{Varol2017,FerraraArming2019} 
is based on a supervised machine learning approach employing Random Forest classifiers~\cite{Breiman2001}. 
Given a Twitter account, Botometer extracts, via the Twitter API, over 1,000 features about the account, 
 including measures of sentiment, time of day, tweets content and Twitter network. 

The immediate output of Botometer is the bot score $S$ ranging over \{0, \ldots 1\}, which however does not represents the probability that the considered account is a bot. The value has to be compared with other scores within a group of accounts, to come up with a plausible ranking.
To interpret the output of Botometer as a bot probability, i.e., to answer the question: `is this account a bot or not?', \mape{the Complete Automation Probability $CAP$ has been introduced in~\cite{FerraraArming2019}:}
%
%
it represents the conditional probability that an account is a bot, given its bot score $S$. 
This conditional probability is calculated by applying the Bayes's rule:  $P(Bot | S) = P(Bot) \frac{P(S|Bot)}{P(S)}$. In~\cite{Varol2017}, an assumption has been made that the prior probability (probability that any randomly-chosen account is  a bot) is $P(Bot)$ = 0.15. However, if one knows the background level of bots in the sample (and this is our case, since the datasets are tagged), $CAP$ can be adjusted with a corrective factor:  
  $CAP* = CAP\frac{P^{d}(Bot)}{0.15}$
 where $P^{d}(Bot)$ is the actual probability that in the domain $d$ under investigation there is a specific number of bots. We know in advance the nature of our datasets, so it is possible to compute $P^{d}(Bot)$.

There are two types of $CAP$, the $CAP\_ENG$ and $CAP\_UNI$: the former exploits textual features regarding tweets in English, the latter  exploits only language-independent features. In the following, we will consider the *version of $CAP\_UNI$, 
that allows for more reliable results even when analysing non-English language tweets.


\subsection{Features related to the account's profile/timeline}\label{subsec:classab}
In a pioneer study about the detection of fake followers~\cite{cresci2015fame}, Cresci et al. tested  Twitter fake followers in the reference set against algorithms based on: (i) classification rules proposed by technology bloggers and social media marketing companies\footnote{Stateofsearch.com: How to Recognize Twitterbots: 7 Signals to Look out for (August 2012) (online newsletter, no more available);
M. Camisani-Calzolari: Analysis of Twitter Followers of the US Presidential Election candidates: Barack Obama and Mitt Romney (August 2012);
Status People Fakers (a former service to tag accounts as fake or not);
SocialBakers (a social media marketing platform, \url{https://www.socialbakers.com/feature/fake-influencers-detection}); 
\url{https://www.twitteraudit.com/}
}, and (ii) feature sets proposed in the literature for detecting spammers. They classified rules and features according to the cost required for gathering the data needed to compute them and showed how the best performing features are also the most costly ones. Then
they implemented a series of lightweight classifiers using less costly features, that nevertheless are 
able to correctly classify more than 95\% of the accounts of the baseline dataset.
Table~\ref{tab:classAB} shows the list of features inherited from~\cite{cresci2015fame} that will be used in the next section to test the outcome of a number of classifiers.
The features in Class A are those obtainable from the account profile data only, 
and those in Class B are those obtainable from the timelines of the accounts. 

\begin{center}
\begin{table}[ht]
\scriptsize
\begin{center}
\begin{tabular}{ll}
    \hline
    {\bf Class A -- profile}&{\bf Class B -- timeline}\\
    \hline
    friends count & rate of posts with at least one Hashtag \\
    followers count & rate of posts with at least one URL\\
    tweets count & rate of posts with at least one mention\\
    $\frac{friends}{followers^2}$ & rate of retweets\\
    account age & \\
    following rate (approximated as $\frac{friends}{age}$) & \\
    the account's profile has a name & \\
    the account's profile has an image & \\
    the account's profile has an address & \\
    the account's profile has a biography & \\
    the account's profile has a URL & \\
    the account belongs to a list & \\
    $2 \times followers \geq friends$ & \\
    $\frac{friends}{followers} \simeq 100$ & \\
    $\frac{friends}{followers} \geq 50$ & \\
    \hline
\end{tabular}
\end{center}
\caption{Class A and Class B features\label{tab:classAB}.}
\end{table}
\end{center}

\subsection{Proportion of tweets sent by unofficial Twitter clients}\label{subsec:makse}
In their study on the dissemination of fake news during the 2016 presidential elections~\cite{bovet2019Nature}, Bovet and Makse proposed to identify tweets originated by bots by inspecting their source field and extracting the name of the Twitter client used to post the tweet. Bovet and Makse considered genuine users only those that mostly tweeted from clients in the list of the official Twitter clients, as listed in the Supplementary Table 14 of~\cite{bovet2019Nature}. The motivation \mape{for not considering
authentic the accounts that tweeted from third-party clients}
is that the latter 
are mainly used by professionals for 
automating some tasks, see, e.g., \url{www.sprinklr.com} or \url{dlvrit.com}.

The approach has been further investigated by the same authors in~\cite{bovet2018validation}, where its classification performances are compared with those of Botometer~\cite{Varol2017}.
The outcomes of the comparison showed a good accuracy of their proposal, although with a higher number of false positives\footnote{False positives are the genuine accounts erroneously classified as bots.} than Botometer. The authors of~\cite{bovet2019Nature,bovet2018validation} also state that evolved bots `might not be detected' by simply looking at the source field, but
`this is also a problem for more advanced methods that rely on a training set of known bots~\cite{Varol2017}'.

In the following, inspired by the evaluation of the source field of tweets, we consider as a feature the proportion of tweets posted by unofficial Twitter clients, on the whole number of tweets of the account. As `whole' number of tweets, we consider the most recent 400 tweets in the account timeline.

\subsection{Experimental Setup}

\begin{table*}[ht]
	\scriptsize
	\centering
	\begin{tabular}{lrr}
 		\toprule
		\textbf{training set} & \#bot & \#human \\
		\midrule
		\texttt{Celebrity-Botwiki} & 642 & 4375 \\
		\texttt{Verified-Botwiki} & 642 & 1919 \\
		\texttt{Verified-Vendor} & 217 & 1919 \\
		\texttt{Stock} & 5619 & 3888 \\
		\texttt{Rtbust}  & 303 & 308 \\
		\bottomrule
	\end{tabular}
	\caption{\small Training sets for the experiments. The accounts are those for which Botometer v3 returns a valid response and that have at least 400 tweets in their timelines.
	\label{tab:training}}
\end{table*}

We combine the datasets described in Section~\ref{sec:datasets} to construct five training sets to test the goodness of the four feature sets in discriminating bots and genuine accounts:
 \texttt{Celebrity}-\texttt{Botwiki}; 
\texttt{Verified}-\texttt{Botwiki}; 
  \texttt{Verified}-\texttt{Vendor};
    \texttt{Stock}; 
 \texttt{Rtbust}. Statistics about the number of bots and genuine accounts in the training sets are in Table~\ref{tab:training}.

The rationale behind the choice of these combinations is to evaluate how the different sets of features are able to recognise recent bot accounts, with more advanced characteristics than bots dated 2010-2017, both for types acting in a coordinated fashion (such as \texttt{Stock} and \texttt{Rtburst}) and for types acting individually (\texttt{Botwiki} and \texttt{Vendor}).

All the training sets do not consider those accounts for which Botometer v3 does not produce a valid response. 
Thus, we will consider  only performance results over the training sets with accounts i) for which Botometer v3 has returned a valid response and ii) have at least 400 tweets in their timeline. 

In the following, we evaluate the performances of five different learning algorithms in terms of well known, standard metrics, such precision, recall, Matthew Correlation Coefficient (MCC) (i.e, the estimator of the correlation between the predicted class and the real class of the samples), and the area-under-the-curve metric (AUC), that is the area under the Receiver Operating Characteristic (ROC) curve~\cite{roc-auc}. As can be seen from the composition of the training sets (Table~\ref{tab:training}), they have different degrees of balance, from \texttt{Rtbust} which is balanced, to \texttt{Verified-Botwiki} and \texttt{Stock} which have a mild degree of imbalance (minority class equal to 20\%-40\% of the whole set), to \texttt{Celebrity-Botwiki} and \texttt{Verified-Vendor} which have a moderate degree of imbalance (minority class equal to 1\%-20\% of the whole set)\footnote{\url{https://developers.google.com/machine-learning/data-prep/construct/sampling-splitting/imbalanced-data}}. 
The ROC-AUC - Area Under the Curve corresponding to the Receiver Operating Characteristics - is generally considered an unbiased metric in imbalanced settings. Provost et al. proposed ROC-AUC as alternatives to accuracy~\cite{weiss2003learning}.
In addition, as evaluation metric, we will also consider the Precision Recall-Area Under the Curve (PR-AUC), often judged less biased than ROC-AUC in the presence of moderate imbalances~\cite{10.1371/journal.pone.0118432}. 
Finally,  we report the Balanced Accuracy, defined as (True Positive rate + True Negative rate)/2.

Concerning the five algorithms, each of them belongs to a different category: MlP (Multilayer Perception)~\cite{159058}, JRip, i.e., a Java-based implementation of the RIPPER algorithm~\cite{COHEN1995115}, Naive Bayes~\cite{10.5555/2074158.2074196}, Random Forest~\cite{rf}, and the Weka~\cite{witten2016data} implementation of the Instance-based Learning Algorithms, i.e., IBk~\cite{ibk}. For all the experiments, we rely on the open source machine learning software Weka. For each evaluation, we apply a 10-fold cross validation on the training sets. For  each algorithm, we use the default parameter settings, detailed in the official Weka documentation\footnote{\url{https://waikato.github.io/weka-wiki/documentation/}}.

As a final note, before presenting the result of our analysis, we would like to highlight that, although we could have directly used the Botometer CAP* to classify accounts as bots or not, we preferred to give the CAP* as a feature to the learning models, exactly as we did with the other features presented. This allows us to give a homogeneous presentation of the results. In the Discussion session, we will show that the two procedures (classifying directly with Botometer/classifying with the learning models) obtain the same performances.



\begin{table*}[t!]
	\scriptsize
	\centering
	\begin{tabular}{lcccccc}
		\toprule
		& \multicolumn{5}{c}{\textbf{Celebrity-Botwiki}} \\
		\cmidrule{2-7}
		Name & Bal. Accuracy & Precision & Recall & MCC & PR-AUC & ROC-AUC \\
\midrule 
\multicolumn{7}{l}{\texttt{Botometer CAP\_UNI$^*$}} \\
MlP & 0.857 & 0.761 & 0.748 & 0.719 & 0.762 & 0.931 \\
JRip & 0.833 & 0.816 & 0.687 & 0.716 & 0.628 & 0.836 \\
NaiveBayes & 0.721 & 0.858 & 0.452 & 0.588 & 0.772 & 0.928 \\
RandomForest & 0.853 & 0.778 & 0.737 & 0.723 & \textbf{0.778} & 0.95 \\
IBk & 0.854 & 0.775 & 0.739 & 0.722 & 0.768 & \textbf{0.951} \\
\midrule 
\multicolumn{7}{l}{\texttt{ClassA}} \\
MlP & 0.961 & 0.925 & 0.932 & 0.918 & 0.978 & 0.996 \\
JRip & 0.984 & 0.983 & 0.971 & 0.974 & 0.97 & 0.984 \\
NaiveBayes & 0.98 & 0.847 & 0.986 & 0.901 & 0.979 & 0.995 \\
RandomForest & 0.989 & 0.991 & 0.979 & 0.983 & \textbf{0.998} & \textbf{1} \\
IBk & 0.951 & 0.912 & 0.915 & 0.901 & 0.843 & 0.955 \\
\midrule 
\multicolumn{7}{l}{\texttt{ClassB}} \\
MlP & 0.954 & 0.914 & 0.921 & 0.905 & 0.95 & 0.976 \\
JRip & 0.962 & 0.922 & 0.935 & 0.918 & 0.92 & 0.966 \\
NaiveBayes & 0.934 & 0.663 & 0.937 & 0.752 & 0.912 & 0.968 \\
RandomForest & 0.964 & 0.941 & 0.937 & 0.93 & \textbf{0.978} & \textbf{0.995} \\
IBk & 0.94 & 0.888 & 0.896 & 0.876 & 0.878 & 0.94 \\
\midrule 
\multicolumn{7}{l}{\texttt{\texttt{Twitter client}}} \\
MlP & 0.962 & 0.804 & 0.958 & 0.859 & \textbf{0.827} & \textbf{0.978} \\
JRip & 0.96 & 0.81 & 0.952 & 0.859 & 0.778 & 0.961 \\
NaiveBayes & 0.951 & 0.614 & 0.993 & 0.743 & 0.796 & 0.977 \\
RandomForest & 0.949 & 0.818 & 0.929 & 0.852 & 0.803 & 0.975 \\
IBk & 0.949 & 0.814 & 0.929 & 0.849 & 0.804 & 0.977 \\
	\bottomrule
	\end{tabular}
	\caption{\small 10-folds cross validation  (imbalanced, \#post$\geqslant$400).}
	\label{tab:botwiki-celebrity}
\end{table*}

\begin{table*}[t!]
	\scriptsize
	\centering
	\begin{tabular}{lcccccc}
		\toprule
		&\multicolumn{5}{c}{\textbf{Verified-Botwiki}} \\
		\cmidrule{2-7}
		Name & Bal. Accuracy & Precision & Recall & MCC & PR-AUC & ROC-AUC \\
\midrule 
\multicolumn{7}{l}{\texttt{Botometer CAP\_UNI$^*$}} \\
MlP & 0.836 & 0.867 & 0.708 & 0.721 & 0.865 & 0.914 \\
JRip & 0.853 & 0.817 & 0.762 & 0.722 & 0.746 & 0.854 \\
NaiveBayes & 0.706 & 0.971 & 0.415 & 0.576 & 0.838 & 0.889 \\
RandomForest & 0.843 & 0.866 & 0.723 & 0.731 & \textbf{0.866} & 0.918 \\
IBk & 0.842 & 0.86 & 0.723 & 0.726 & \textbf{0.866} & \textbf{0.919} \\
\midrule 
\multicolumn{7}{l}{\texttt{ClassA}} \\
MlP & 0.958 & 0.909 & 0.948 & 0.903 & 0.948 & 0.986 \\
JRip & 0.976 & 0.975 & 0.96 & 0.957 & 0.958 & 0.974 \\
NaiveBayes & 0.961 & 0.907 & 0.955 & 0.907 & 0.964 & 0.982 \\
RandomForest & 0.983 & 0.989 & 0.969 & 0.972 & \textbf{0.997} & \textbf{0.999} \\
IBk & 0.952 & 0.908 & 0.935 & 0.895 & 0.866 & 0.951 \\
\midrule 
\multicolumn{7}{l}{\texttt{ClassB}} \\
MlP & 0.962 & 0.965 & 0.935 & 0.933 & 0.963 & 0.966 \\
JRip & 0.982 & 0.958 & 0.979 & 0.957 & 0.953 & 0.986 \\
NaiveBayes & 0.941 & 0.823 & 0.951 & 0.843 & 0.951 & 0.979 \\
RandomForest & 0.983 & 0.976 & 0.974 & 0.966 & \textbf{0.993} & \textbf{0.997} \\
IBk & 0.973 & 0.964 & 0.958 & 0.948 & 0.957 & 0.977 \\
\midrule 
\multicolumn{7}{l}{\texttt{\texttt{Twitter client}}} \\
MlP & 0.958 & 0.867 & 0.965 & 0.885 & \textbf{0.878} & \textbf{0.973} \\
JRip & 0.958 & 0.884 & 0.957 & 0.892 & 0.833 & 0.949 \\
NaiveBayes & 0.934 & 0.728 & 0.993 & 0.793 & 0.857 & 0.967 \\
RandomForest & 0.956 & 0.878 & 0.957 & 0.887 & 0.874 & \textbf{0.973} \\
IBk & 0.956 & 0.874 & 0.957 & 0.884 & 0.874 & \textbf{0.973} \\
	\bottomrule
	\end{tabular}
	\caption{\small 10-folds cross validation  (imbalanced, \#post$\geqslant$400).}
	\label{tab:botwiki-verified}
\end{table*}

\section{Experimental results}\label{sec:exp}
In this section, we report our experiments and comment their main outcomes.

\subsection{Celebrity-Botwiki}
Table~\ref{tab:botwiki-celebrity} shows the performances results of the same learning algorithms when fed with the Botometer v3 CAP\_UNI*, the Class A and Class B features, and the percentage of tweets posted by unofficial clients, for the training set composed of self-declared bots and the celebrity accounts.

All the four approaches achieved very good results in terms of ROC-AUC. It can be seen that the other metrics are slightly worse for models trained with the CAP\_UNI* score returned by Botometer v3. To be precise, the worst results are those obtained with the CAP\_UNI* and with the feature about the Twitter client. The performance with the latter is indeed not so surprising, since celebrity accounts are probably managed by social media managers who may also decide to automate the tweets posted. 
Regarding the CAP\_UNI*, for this particular training set some celebrities have been tagged as bots (lower precision), and some bots have been classified as celebrities (lower recall). This also results in slightly lower values for the ROC-AUC. As one of the most skewed training sets, we notice a worsening of the PR-AUC compared to the ROC-AUC for every tested feature set, but Class A and Class B still give values greater than 0.9. 
\begin{table*}[t!]
	\scriptsize
	\centering
	\begin{tabular}{lcccccc}
		\toprule
		& \multicolumn{5}{c}{\textbf{Verified-Vendor}} \\
		\cmidrule{2-7}
		Name & Bal. Accuracy & Precision & Recall & MCC & PR-AUC & ROC-AUC \\
\midrule 
\multicolumn{7}{l}{\texttt{Botometer CAP\_UNI$^*$}} \\
MlP & 0.764 & 0.879 & 0.535 & 0.661 & 0.689 & 0.854 \\
JRip & 0.771 & 0.84 & 0.553 & 0.654 & 0.551 & 0.75 \\
NaiveBayes & 0.709 & 0.948 & 0.42 & 0.608 & 0.681 & 0.838 \\
RandomForest & 0.748 & 0.84 & 0.507 & 0.625 & 0.707 & 0.885 \\
IBk & 0.767 & 0.843 & 0.544 & 0.65 & \textbf{0.709} & \textbf{0.887} \\
\midrule 
\multicolumn{7}{l}{\texttt{ClassA}} \\
MlP & 0.767 & 0.799 & 0.549 & 0.632 & 0.711 & 0.862 \\
JRip & 0.848 & 0.812 & 0.715 & 0.737 & 0.649 & 0.828 \\
NaiveBayes & 0.781 & 0.868 & 0.572 & 0.679 & 0.739 & 0.906 \\
RandomForest & 0.858 & 0.859 & 0.729 & 0.77 & \textbf{0.837} & \textbf{0.939} \\
IBk & 0.75 & 0.634 & 0.535 & 0.54 & 0.399 & 0.75 \\
\midrule 
\multicolumn{7}{l}{\texttt{ClassB}} \\
MlP & 0.816 & 0.88 & 0.641 & 0.728 & 0.767 & 0.913 \\
JRip & 0.814 & 0.767 & 0.65 & 0.676 & 0.629 & 0.832 \\
NaiveBayes & 0.605 & 0.598 & 0.226 & 0.329 & 0.518 & 0.901 \\
RandomForest & 0.836 & 0.828 & 0.687 & 0.73 & \textbf{0.819} & \textbf{0.935} \\
IBk & 0.804 & 0.745 & 0.632 & 0.654 & 0.517 & 0.815 \\
\midrule 
\multicolumn{7}{l}{\texttt{\texttt{Twitter client}}} \\
MlP & 0.5 & 0 & 0 & 0 & \textbf{0.146} & 0.58 \\
JRip & 0.5 & 0 & 0 & 0 & 0.101 & 0.495 \\
NaiveBayes & 0.5 & 0 & 0 & 0 & 0.136 & \textbf{0.607} \\
RandomForest & 0.511 & 0.242 & 0.033 & 0.055 & 0.135 & 0.6 \\
IBk & 0.5 & 0.1 & 0.028 & 0 & 0.133 & 0.605 \\

	\bottomrule
	\end{tabular}
	\caption{\small 10-folds cross validation  (imbalanced, \#post$\geqslant$400).}
	\label{tab:vendor-verified}
\end{table*}

\begin{table*}[t!]
	\scriptsize
	\centering
	\begin{tabular}{lcc}
		\toprule
		Name & 10\%-90\% & 40\%-60\% \\
\midrule 
\multicolumn{2}{l}{\texttt{Botometer CAP\_UNI$^*$}} \\
MlP & 0.535 & 0.683 \\
JRip & 0.553 & 0.692 \\
NaiveBayes & 0.42 & 0.563 \\
RandomForest & 0.507 & 0.673 \\
IBk & 0.544 & 0.692 \\
\midrule 
\multicolumn{2}{l}{\texttt{ClassA}} \\
MlP & 0.549 & 0.756 \\
JRip & 0.715 & 0.848 \\
NaiveBayes & 0.572 & 0.6 \\
RandomForest & 0.729 & 0.853 \\
IBk & 0.535 & 0.738 \\
\midrule 
\multicolumn{2}{l}{\texttt{ClassB}} \\
MlP & 0.641 & 0.807 \\
JRip & 0.65 & 0.779 \\
NaiveBayes & 0.226 & 0.738 \\
RandomForest & 0.687 & 0.807 \\
IBk & 0.632 & 0.816 \\

	\bottomrule
	\end{tabular}
	\caption{\small Vendor-Verified training set: Recall values for the original set and for a modified set where the balance among the classes is 40\% Vendor - 60\% Verified\label{tab:vendor-verified-recall}}
\end{table*}

\subsection {Verified-Botwiki}
Table~\ref{tab:botwiki-verified} shows that, when the training set is composed by Botwiki and the verified Twitter accounts, the outcome is very similar to the previous one. 
\begin{table*}[t!]
	\scriptsize
	\centering
	\begin{tabular}{lcccccc}
		\toprule
		& \multicolumn{5}{c}{\textbf{Stock}} \\
		\cmidrule{2-7}
		Name & Bal. Accuracy & Precision & Recall & MCC & PR-AUC & ROC-AUC \\
\midrule 
\multicolumn{7}{l}{\texttt{Botometer CAP\_UNI$^*$}} \\
MlP & 0.797 & 0.815 & 0.886 & 0.609 & 0.826 & 0.823 \\
JRip & 0.797 & 0.812 & 0.893 & 0.611 & 0.79 & 0.789 \\
NaiveBayes & 0.787 & 0.839 & 0.795 & 0.569 & 0.829 & 0.834 \\
RandomForest & 0.794 & 0.811 & 0.887 & 0.605 & \textbf{0.838} & \textbf{0.839} \\
IBk & 0.794 & 0.81 & 0.89 & 0.604 & 0.837 & \textbf{0.839} \\
\midrule 
\multicolumn{7}{l}{\texttt{ClassA}} \\
MlP & 0.717 & 0.762 & 0.793 & 0.438 & 0.822 & 0.787 \\
JRip & 0.795 & 0.829 & 0.842 & 0.592 & 0.818 & 0.814 \\
NaiveBayes & 0.561 & 0.623 & 0.981 & 0.235 & 0.85 & 0.804 \\
RandomForest & 0.801 & 0.835 & 0.842 & 0.603 & \textbf{0.914} & \textbf{0.882} \\
IBk & 0.674 & 0.732 & 0.74 & 0.348 & 0.7 & 0.673 \\
\midrule 
\multicolumn{7}{l}{\texttt{ClassB}} \\
MlP & 0.824 & 0.834 & 0.909 & 0.664 & \textbf{0.872} & \textbf{0.869} \\
JRip & 0.825 & 0.836 & 0.907 & 0.666 & 0.825 & 0.835 \\
NaiveBayes & 0.806 & 0.82 & 0.898 & 0.63 & 0.852 & 0.854 \\
RandomForest & 0.806 & 0.827 & 0.875 & 0.62 & 0.871 & 0.863 \\
IBk & 0.753 & 0.795 & 0.807 & 0.507 & 0.782 & 0.763 \\
\midrule 
\multicolumn{7}{l}{\texttt{\texttt{Twitter client}}} \\
MlP & 0.605 & 0.661 & 0.815 & 0.231 & \textbf{0.674} & 0.635 \\
JRip & 0.64 & 0.688 & 0.819 & 0.302 & \textbf{0.674} & 0.641 \\
NaiveBayes & 0.558 & 0.622 & 0.96 & 0.202 & 0.661 & 0.62 \\
RandomForest & 0.615 & 0.663 & 0.874 & 0.273 & 0.671 & \textbf{0.643} \\
IBk & 0.607 & 0.657 & 0.879 & 0.258 & 0.668 & 0.64 \\
	\bottomrule
	\end{tabular}
	\caption{\small 10-folds cross validation  (imbalanced, \#post$\geqslant$400).}
	\label{tab:stock}
\end{table*}
\subsection{Verified-Vendor}\label{subsec:vendor-verified} When we consider the dataset formed by verified accounts and fake followers, the results of the various approaches start to differ significantly, see Table~\ref{tab:vendor-verified}. The Twitter client approach fails on every front. 
This training set  is the most unbalanced of all, in fact fake followers represent only 10\% of the whole set. To limit the perhaps too optimistic view given by the ROC-AUC, we can analyze the PR-AUC values, which are derived from the individual precision and recall values. Leaving aside Naive Bayes and IBk, which for class A and B give much worse results than CAP*, the precision for the other types of algorithms is comparable for the three set of features. Recall gets the highest value for Class A features and Random Forest (0.729). However, for all the feature sets, the algorithms classify many fake followers as verified accounts. 

\mape{We have examined this aspect in more detail: the low recall values could in fact be due to the strong unbalance of the dataset. Table~\ref{tab:vendor-verified-recall} compares, for each algorithm, the recall calculated on the unbalanced dataset and the recall calculated on the dataset where the majority class (the verified accounts) has been under-sampled\footnote{\url{https://weka.sourceforge.io/doc.dev/weka/filters/supervised/instance/SpreadSubsample.html}}. In this way, the dataset is composed of 40\% Vendor accounts and 60\% Verified accounts. The recall values in the balanced dataset actually increase. Even if they do not reach the peaks of 0.9 of the datasets we have analysed so far, they still oscillate between 0.6 and 0.853.
We will come back to this aspect in Section~\ref{sec:findings} of the Discussion section.
}



\subsection{Stock}
We now analyse a dataset consisting of both bots and genuine accounts, each of which tweet about financial markets. Some of the bots act in a coordinated manner. The dataset is very recent and this let us assume that we are dealing with a very sophisticated version of bots. First, from Table~\ref{tab:stock}, we can see that the values of ROC-AUC and PR-AUC are similar. The dataset is in fact almost balanced, bots are about 60\% of the total accounts in the dataset. Again, analysing the Twitter client does not yield good results (although this approach performs better than in the case considered in Section~\ref{subsec:vendor-verified}). Surprisingly, the remaining three approaches  have similar results and they generally obtain values greater than 0.8 for ROC-AUC (11 times out of 15). Class A generally obtains the lowest values for precision and recall, while class B gives the best results. We can conjecture that the nature of these bots, which is peculiar to the content tweeted, means that the timeline features of the accounts are sufficient, and necessary, to best detect these accounts. We would like to point out that in the paper where \mape{these accounts} were first analysed~\cite{crescirtbust2019}, the proposed detection technique is an unsupervised technique that detects these bot accounts as a group acting in a coordinated manner. Our analysis suggests that we could obtain a good result at detection level, even leveraging a traditional supervised classifier, exploiting features that are not expensive in terms of data gathering.
\begin{table*}[t!]
	\scriptsize
	\centering
	\begin{tabular}{lcccccc}
		\toprule
		& \multicolumn{5}{c}{\textbf{Rtbust}} \\
		\cmidrule{2-7}
		Name & Bal. Accuracy & Precision & Recall & MCC & PR-AUC & ROC-AUC \\
\midrule 
\multicolumn{7}{l}{\texttt{Botometer CAP\_UNI$^*$}} \\
MlP & 0.538 & 0.53 & 0.588 & 0.075 & 0.574 & 0.588 \\
JRip & 0.616 & 0.612 & 0.614 & 0.231 & 0.552 & \textbf{0.592} \\
NaiveBayes & 0.529 & 0.614 & 0.152 & 0.088 & 0.554 & 0.554 \\
RandomForest & 0.562 & 0.56 & 0.542 & 0.123 & 0.573 & 0.571 \\
IBk & 0.559 & 0.55 & 0.604 & 0.118 & \textbf{0.585} & 0.569 \\
\midrule 
\multicolumn{7}{l}{\texttt{ClassA}} \\
MlP & 0.717 & 0.767 & 0.618 & 0.442 & 0.761 & 0.764 \\
JRip & 0.738 & 0.77 & 0.674 & 0.48 & 0.737 & 0.746 \\
NaiveBayes & 0.649 & 0.6 & 0.865 & 0.329 & 0.7 & 0.742 \\
RandomForest & 0.776 & 0.801 & 0.73 & 0.554 & \textbf{0.867} & \textbf{0.841} \\
IBk & 0.706 & 0.698 & 0.717 & 0.412 & 0.644 & 0.706 \\
\midrule 
\multicolumn{7}{l}{\texttt{ClassB}} \\
MlP & 0.827 & 0.869 & 0.766 & 0.658 & 0.871 & \textbf{0.877} \\
JRip & 0.805 & 0.852 & 0.736 & 0.616 & 0.815 & 0.827 \\
NaiveBayes & 0.747 & 0.711 & 0.826 & 0.5 & 0.816 & 0.832 \\
RandomForest & 0.818 & 0.859 & 0.76 & 0.641 & \textbf{0.902} & 0.869 \\
IBk & 0.778 & 0.769 & 0.789 & 0.556 & 0.707 & 0.77 \\
\midrule 
\multicolumn{7}{l}{\texttt{\texttt{Twitter client}}} \\
MlP & 0.514 & 0.51 & 0.522 & 0.028 & 0.524 & 0.539 \\
JRip & 0.596 & 0.57 & 0.743 & 0.2 & \textbf{0.532} & \textbf{0.564} \\
NaiveBayes & 0.537 & 0.517 & 0.921 & 0.115 & 0.52 & 0.537 \\
RandomForest & 0.564 & 0.539 & 0.802 & 0.145 & 0.51 & 0.543 \\
IBk & 0.575 & 0.546 & 0.832 & 0.175 & 0.52 & 0.555 \\

	\bottomrule
	\end{tabular}
	\caption{\small 10-folds cross validation  (\#post$\geqslant$400).}
	\label{tab:rtbust}
\end{table*}

\subsection{Rtbust}
We now analyse the classification results on the last dataset, consisting of a mix of genuine accounts and bots. The latter feature a coordinated behaviour, whose peculiarity is to retweet {\it en masse} some target account. 
The dataset is perfectly balanced (303-308), so we focus on ROC-AUC rather than PR-AUC. The result is that, once again, Class B gives the best results, reaching values above 0.8 and peaking at 0.877. The Class A approach degrades the performances; 0.841 is the best result, but with all other values between 0.7 and 0.8. We argue that this is a surprising result, for the following reasons. First, Class B features are not onerous to compute compared to the ensemble of features used in Botometer v3 (in our experiments summarised by CAP\_UNI*). Secondly,  the authors  that firstly discovered these kind of bots achieved, in their best parameters' configuration,  a precision of 0.934 (w.r.t. our best result = 0.869), a recall of 0.814, comparable to our recall value of 0.826, and MCC = 0.757 (w.r.t. our best result = 0.658)~\cite{crescirtbust2019}. Although achieving lower performances than~\cite{crescirtbust2019}, we obtain promising results without relying on a specialised technique aimed at detecting retweeters. It is also worth noting that Cresci et al.~\cite{crescirtbust2019} experimented with different parameters' configurations, and, except for their values of 0.757, the MCCs values that we obtain here are always higher  than those obtained in~\cite{crescirtbust2019} for  the tested configurations.
 Last, but not least, our approach and its outcome align us with the very recent approach in~\cite{DBLP:conf/cikm/Sayyadiharikandeh20},  without leveraging the premium version of Botometer v4 (see a discussion in the following section).

\section{Discussion}\label{sec:discussion}
\mape{In this section, we 1) highlight our motivations for considering features easily obtainable from the account profile and timeline, through an analysis of the cost of data gathering from Twitter; 2) discuss the main findings of the manuscript; 3) introduce some limitations of the work and test its robustness by introducing further experiments.}

\subsection{\mape{Twitter API and Botometer rate limits, in a nutshell}
}
In this paper, we have carried out some experiments  over datasets of genuine and bot accounts, in order to compare the performances of well-known learning algorithms
used to detect bots, while considering  different sets of state-of-art features of the accounts.
From the simplest to the most complex ones, the account features are computed from the data that Twitter makes available from the APIs\footnote{Twitter API documentation: \url{https://developer.twitter.com/en/docs/twitter-api}}. As originally defined by Cresci et al. in~\cite{cresci2015fame}, such features can be distinguished, according to the number of API calls needed to compute them, into features related to the account profile (so called Class A features), related to the timeline (Class B features), and related to relationships (Class C features). Extracting features other than those that can be computed from data in the account profile have a higher cost in terms of the number of API calls needed.

To retrieve the most recent tweets posted by one user, Twitter provides two versions of the same service, namely v1 and v2. At time of writing, the Twitter documentation reports that the new API v2 will fully replace the v1.1 standard, premium, and enterprise APIs. 

The user timeline API v2
limits the number of tweets that can be obtained per request, compared to v1:
the latter allows up to a maximum of 200 tweets per single request, while the former allows up to a maximum of 100 tweets. Moreover, the maximum number of requests is equal to 1500 requests per 15-minute window.

Similarly, to get the list of IDs of followers and followings of an account, there are new versions of the APIs.
With respect to v1, the number of IDs per request has been greatly reduced, from 5,000 in v1 to 1,000 in v2.
Concerning the rate limit, v2 allows for 15 requests per 15-minute window. 
Thus, if for example we concentrate on the account’s followers, downloading the whole list of
followers of an account would require $\left\lceil\frac{{\it f}}{5000}\right\rceil$ API calls, where $f$ is the number of the account's followers. 
The number of followers and followings of an account is not limited and, therefore, it is impossible to calculate the maximum number of API calls to crawl the necessary information and compute the features related to the relationships of an account. As an example, the Twitter account with the highest number of followers belonged, at time of writing, to Barack Obama (\textit{@BarackObama}), with around 130 millions accounts. To gather information about them,  one would need around 130,000 Twitter API calls. 

In light of the above, it is clearly more appealing to use `low cost' features, to reduce the time for feature gathering. Hence the idea of the manuscript, i.e., to test efficacy of low cost features on more advanced bot datasets.

A tool such as Botometer can save the user the process of feature extraction - leaving this part of the work to Botometer itself. In this regard,
we would like to clarify that our goal here is not to evaluate or question the quality of a system like Botometer. This well-known bot detector is now at version 4, and it has been recently shown  that it performs well for  detecting both single-acting bots and coordinated campaigns~\cite{DBLP:conf/cikm/Sayyadiharikandeh20}. 
Botometer has however some limitations that may justify considering alternative approaches for bot detection.  
For much of its life, Botometer was an openly accessible tool, relying on the Twitter API to collect recent data about the accounts to investigate. As highlighted above, in that case the query limits imposed by the Twitter API caused the analysis of large population of accounts to be very time consuming. 
Botometer v3 imposes additional rate limits, while the current version v4
provides 
 both a free version and a premium version\footnote{\url{https://cnets.indiana.edu/blog/2020/09/01/botometer-v4/}}.  The premium version (50 dollars/month) allows a rate limit of 17,280 requests per day (each request processes 1 user only).  
In addition, v4 premium offers the lite version BotometerLite, which does not interface with Twitter, but simply takes the tweet, retrieves the author, and does the necessary follow-up analysis. 
Interestingly, this light version only needs the information in the user profile to perform bot detection. This resembles the approach considered in this paper, the one based on Class A features. The disadvantage is that each request to BotometerLite can process a maximum of 100 users,  with the limit of 200 requests per day, leading to a maximum of 20k account checks per day.
%

\subsection{\mape{Classifying via CAP* as output of Botometer or as input to the learning models}}
\begin{table*}[ht]
	\scriptsize
	\centering
	\begin{tabular}{llccc}
 		\toprule
		\textbf{Model} & \textbf{Thd} & Bal. Accuracy & Precision & Recall \\
		\midrule
		\multicolumn{4}{l}{\texttt{Celebrity-Botwiki}} \\
        Rule-based & 0.015 & 0.855 & 0.773 & 0.742\\
        MlP & - & 0.857 & 0.761 & 0.748\\
        \midrule 
        \multicolumn{4}{l}{\texttt{Verified-Botwiki}} \\
        Rule-based & 0.025 & 0.858 & 0.829 & 0.768\\
        JRip & - & 0.853 & 0.817 & 0.762\\
        \midrule 
        \multicolumn{4}{l}{\texttt{Verified-Vendor}} \\
        Rule-based & 0.025 & 0.779 & 0.822 & 0.572\\
        JRip & - & 0.771 & 0.84 & 0.553\\
        \midrule 
        \multicolumn{4}{l}{\texttt{Stock}} \\
        Rule-based & 0.4 & 0.8 & 0.824 & 0.869\\
        RandomForest & - & 0.794 & 0.811 & 0.887\\
        \midrule 
        \multicolumn{4}{l}{\texttt{RtBurst}} \\
        Rule-based & 0.02 & 0.624 & 0.615 & 0.647\\
        JRip & - & 0.616 & 0.612 & 0.614\\
		\bottomrule
	\end{tabular}
	\caption{Performances of Botometer and our classifiers that use the Botometer's CAP* as their feature
	\label{tab:Botometer-models-comparison}}
\end{table*}

For the sake of completeness, here we compare the performance of Botometer, based directly on CAP, and of our classifiers that use CAP as their input feature. As stated in Section~\ref{sec:botometer}, we consider here the *version of CAP\_UNI, i.e., CAP\_UNI adjusted with a corrective factor taking into account the actual number of bots in the training sets. 
Once having CAP\_UNI* for every account in the training sets, we measured the Botometer’s performances by applying a threshold-based rule: if CAP\_UNI* $\ge th$, then the account is tagged as a bot. Then, we evaluated the performances of the rule varying the threshold in the interval [0, \ldots 1]. Finally, for each training set, we selected the threshold that gave the best Balanced Accuracy. This value of Balanced Accuracy (and the related Precision and Recall values) have been compared with the best performances obtained by our learning models. The results of the comparison are in Table~\ref{tab:Botometer-models-comparison}. For each training set, we have reported the Botometer threshold and the learning model for which the best Balanced Accuracy was obtained. 
We have also reported the corresponding Precision and Recall values. 
As the results show, there is no substantial difference in considering CAP\_UNI* directly, or in giving it as input to the learning models. 
This gives us confidence that the choice we made is trustworthy. This way, we could maintain uniformity of both the adopted methodology and the presentation of results (Section~\ref{sec:exp}), i.e., always considering feature sets + learning models.

\subsection{\mape{Main findings}}\label{sec:findings} From the study conducted in this paper, we can draw the following findings. 
\begin{itemize}
\item Basic methods based on a single, basic, feature, like the Twitter client, are not sufficient to detect sophisticated bots, possibly acting in a coordinated fashion. For the specific training sets considered here, the method gives good results only when considering 
rather simplistic bots that act individually, i.e., \texttt{Botwiki}. It is interesting instead to note the low performances obtained considering the Twitter client with regard to \texttt{Vendor}: Although they act individually, the Twitter client is not enough to recognise them, because they do not tweet by third-parties clients;
\item Surprisingly, a set of simple features, such as those obtained from the user's timeline, are effective in distinguishing novel social bots, either acting individually (like those in the \texttt{Vendor} dataset) or in teams (like the retweeters in \texttt{Rtbust} and the promoters of low-value stocks in \texttt{Stock}). 
Obviously, deeming an individual account as a bot does not imply that it is part of a team programmed for the same goal. 
We think, however, that an approach based on features such as those in Class A and Class B, once proven to work well on datasets in which we know there are bot squads -such as those investigated in this manuscript- can be useful as a first step for finding such squads.
Then, the nature of the unveiled bots 
can be further investigated by exploiting specialised detection systems. 
As a notable example, the Social Fingerprinting bot detection technique~\cite{cresci2018social} 
encodes the online behavior of a user into a string of characters that represents its digital DNA. Then, digital DNA sequences are compared between one another by means of  bioinformatics algorithms. The technique classifies as bots acting in teams those users that have suspiciously high similarities among their digital DNA sequences.
To obtain the digital DNA of the user, various encodings can be used. If we aim to understand, for example, if within a set of bots there are some that advertise the same product, or point to the same piece of information or account, we could think about coding their Twitter timelines according to the nature of hashtags, mentions and URLs therein.
Some recent work did something similar, even if not under the flag of digital DNA, see, e.g., 
~\cite{caldarelli2020immi,DBLP:conf/icwsm/HuiYTM20};
\item When considering the analyses launched on the \texttt{Vendor-Verified} set, all the tested features lead to a collapse of the recall values. \mape{In Table~\ref{tab:vendor-verified-recall}, we have verified that the recall rises when the training set is more balanced. In fact, even if the values do not reach peaks of 0.9, they are similar to the values obtained considering the training set \texttt{Rtbust}, see Table~\ref{tab:rtbust}.} Thus, compared to the other sets (\texttt{Celebrity-Botwiki}, \texttt{Verified-Botkiwi}, \texttt{Stock}, some
 fake followers in the \texttt{Vendor} set are erroneously tagged as genuine Twitter accounts. 
This suggests that even this strain of account, originally designed in a very simplistic way, just to increase the number of followers of a Twitter account,  has evolved. 
\end{itemize}


\subsection{\mape{Considering timelines shorter than 400 tweets}} We are aware that our study has limitations. First, in order to have a solid comparison with similar approaches, like the one proposed in~\cite{DBLP:conf/cikm/Sayyadiharikandeh20}, we would have to test our feature sets on the datasets used in that work. This is far from impossible, as the datasets are known: we put it on the agenda for future work. 

\begin{figure}[ht!]
\centering
\includegraphics[width=.47\linewidth]{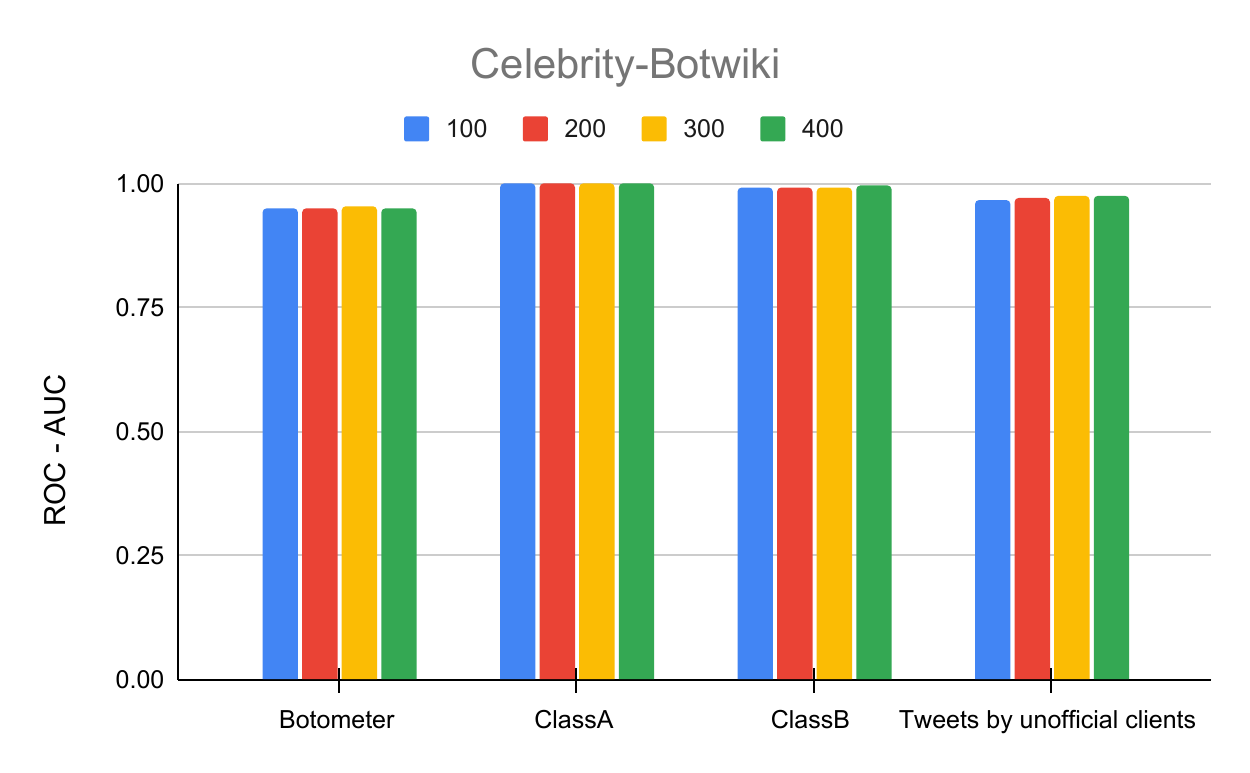} 
\includegraphics[width=.47\linewidth]{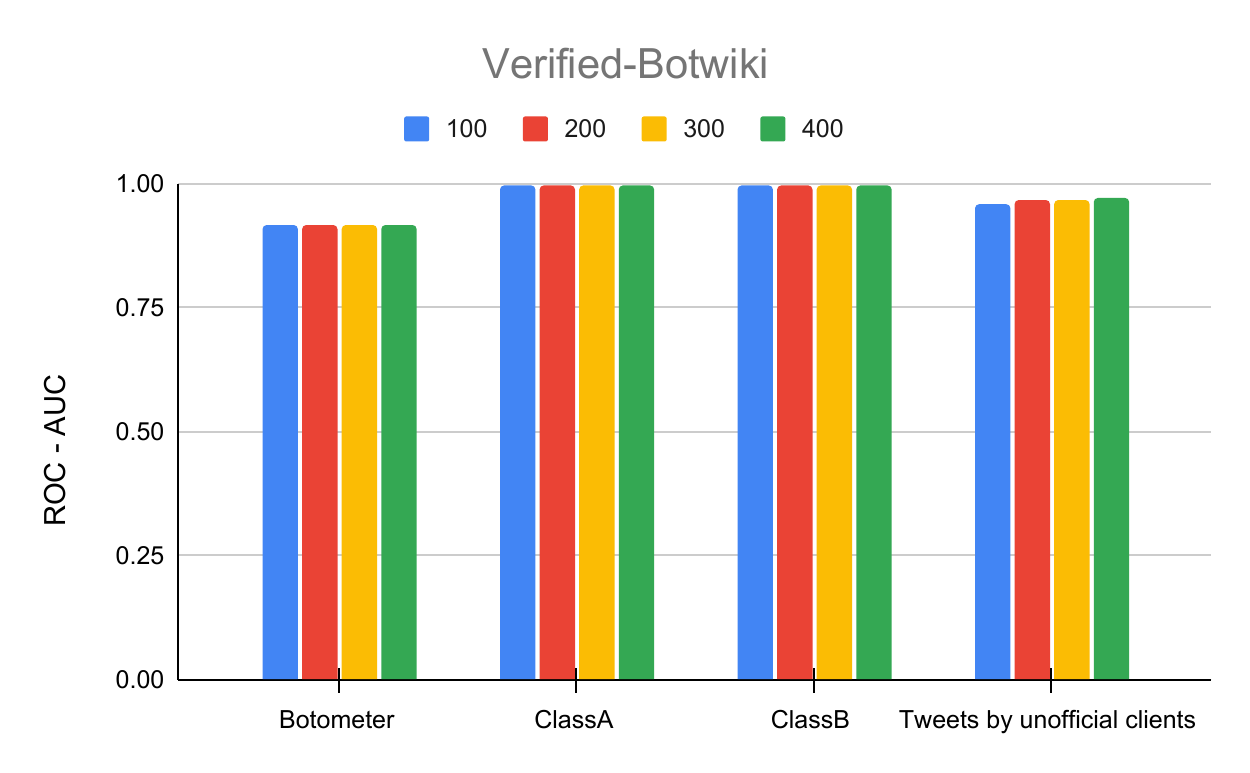}
\includegraphics[width=.47\linewidth]{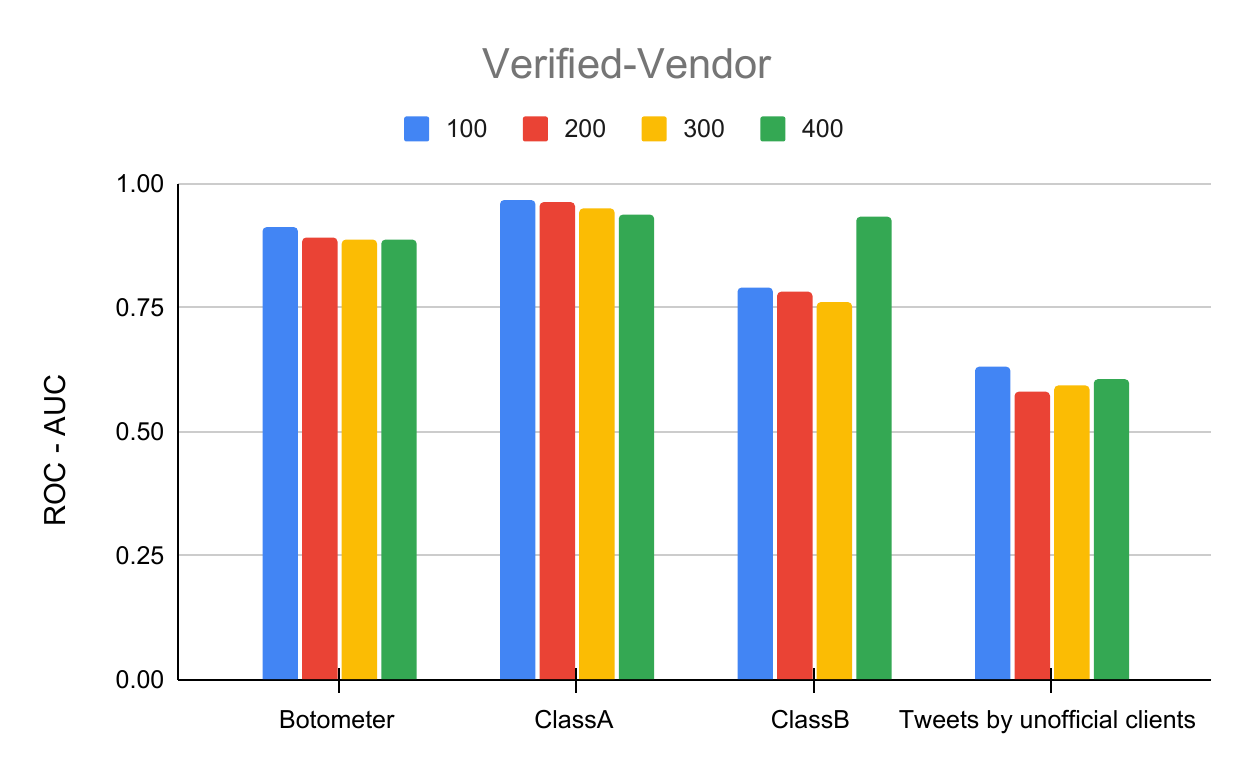}
\includegraphics[width=.47\linewidth]{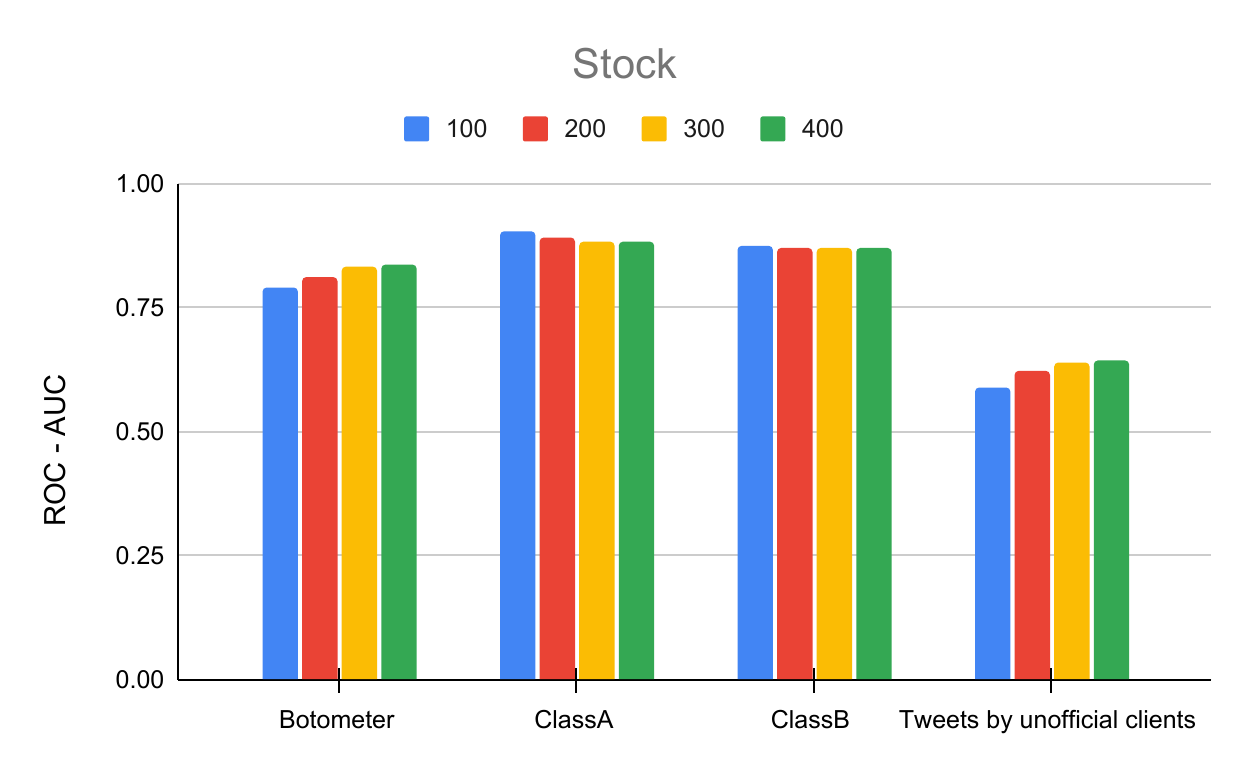}
\includegraphics[width=.47\linewidth]{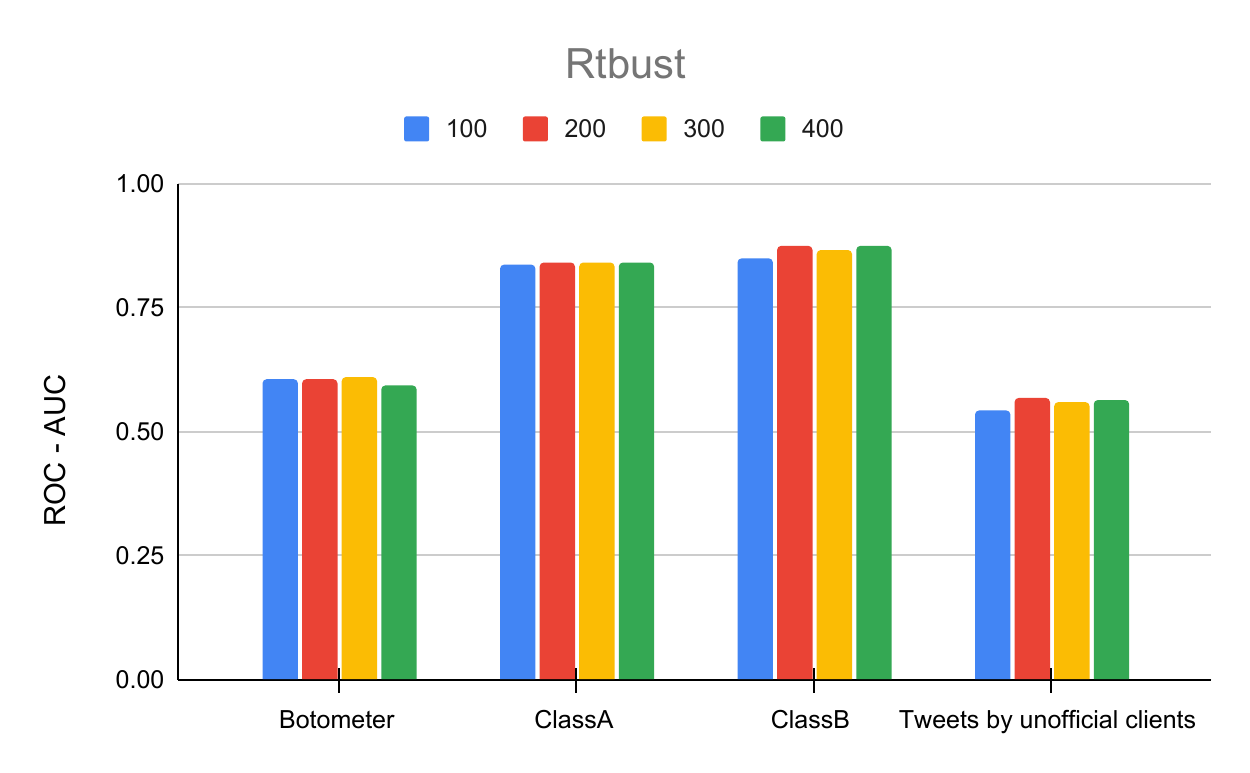}
    \caption{ROC-AUC by varying the minimum number of posts in the account timelines, per training set. \label{fig:timeline_lenght_inv}}
\end{figure}

Then, we guessed about a possible change in the performance results by considering timelines shorter than 400. In fact, as can be seen from Table~\ref{tab:actual-datasets}, the number of accounts in the datasets changes if we set thresholds on the minimum number of tweets that each account must have. For example, considering the \texttt{Vendor} dataset, there is a net decrease for those accounts that have at least 400 tweets (217 accounts) compared to the accounts currently `alive' at the time of data re-crawling (699 accounts). 
Therefore, we have analysed how the ROC-AUC varies with respect to a variation of the number of posts in the accounts' timeline\footnote{\mape{In this experiment, we do not show the PR-AUC, given the fact that this metric `should not be compared between populations' with different bot accounts prevalence, `because their values are prevalence dependent'~\cite{OZENNE2015855}.}}. Considering only accounts that have at least {\it x} tweets in their timeline leads to a change of the training sets (see Table~\ref{tab:actual-datasets}). In order to demonstrate the robustness of our analysis, we should test if the ROC-AUC remains constant on the training sets, when varying the minimum number that an account must have in its timeline.
As can be seen from Figure~\ref{fig:timeline_lenght_inv}, the ROC-AUC values remain stable, except in the case of the \texttt{Verified-Vendor} training set, which shows a marked difference between the AUC values for 100, 200, and 300 tweets (0.73, 0.73, 0.75 respectively) and the value for 400 tweets (>0.9) for the Class B feature set. In this case, the performance degrades when fewer tweets are considered. 

\subsection{\mape{Feature analysis}}

\begin{table*}[ht]
	\scriptsize
	\centering
	\begin{tabular}{ll}
 		\toprule
		\textbf{Verified-Botwiki} & \textbf{Stock} \\
		\midrule
(1.000) rate of retweets & (1.000) rate of retweets\\
(0.955) rate of posts with at least one mention & (0.942) rate of posts with at least one Hashtag\\
(0.928) friends count & (0.925) Botometer\\
(0.907) rate of posts with at least one Hashtag & (0.865) followers count\\
(0.897) Twitter client & (0.793) rate of posts with at least one URL\\
		\bottomrule
	\end{tabular}
	\caption{Ranking of relevance of the features 
	\label{tab:infogain}}
\end{table*}

Here, we show  the results on the relevance of the single features, based on the Information Gain, a measure about the informativeness of a feature with respect to the predicting class~\cite{kent1983information}. The information gain  can be informally defined as the expected reduction in entropy caused by the knowledge of the value of a given attribute~\cite{mitchell1997machine}. The analysis has been implemented in  Weka, through the attribute selector algorithm \textit{InfoGainAttributeEval}\footnote{InfoGainAttributeEval: \url{https://tinyurl.com/ve99qt8}}.
In Table~\ref{tab:infogain}, we report the ranking of the first five most important features, for two specific training sets, according to the Information Gain value. We show  the results obtained for one of the `easiest-to-classify' training set, namely the \texttt{Verified-Botwiki}, composed by simplistic bots and verified Twitter users, and one of the most `troublesome-to-classify' one, where the bots work in team, i.e., \texttt{Stock}. 
Not surprisingly, the Twitter client ranks 5th amongst the most relevant features for the \texttt{Verified-Botwiki} training set, while, looking at the \texttt{Stock} training set, the same feature does not appear in the top 5 results, as shown in the table, but not even in the top 12 (actually, it ranks 13th). 
Looking instead at the results for \texttt{Stock}, we can see that in the first five positions there are class A and B features and the CAP*. We consider this an interesting outcome, which would confirm that features and tools coined a few years ago are still effective for the detection of novel social bots.




\section{Conclusions}\label{sec:concl}
In this paper, we carried out an exploratory study on the effectiveness of different set of features exploitable for bot detection methods. In particular, we focused on features that, for the complexity of their computation, require minimal analytical efforts. 
The considered features are well known by Academia and commonly exploited in the past for bot detection. In particular, Class A and Class B feature sets were first introduced in~\cite{cresci2015fame},  limited to the detection of Twitter fake followers, and have given excellent performance results. The Botometer scores have been judged more than effective indicators for automated account detection for years. Finally, the idea of using the source field of the tweet to evaluate from which client one tweets was proposed in~\cite{bovet2019Nature}. 
\mape{We examined such existing features
on new bot designs to identify those feature sets that allow us to build  bot detectors that are efficient and
effective, also for unveiling  newer and more sophisticated accounts.}


We have learned the following main lessons: 
\begin{enumerate}
\item The method based on considering the Twitter client has a minimal complexity, but it proves insufficient to detect sophisticated bots; 
\item Profile and -especially- timeline features prove to be significant in discriminating even advanced bots, i.e., those working in teams;
\item The evolution of bots does not stop: fake followers, a typology of bots that was once very simplistic, now manage to well disguise themselves amongst the genuine accounts.
\end{enumerate}

The experiments shown in this article are not exhaustive. More accurate analyses can be carried out considering, e.g.,  a different composition of the training sets analysed and other datasets. 
We believe, however, that the results shown in this paper, and the related discussions, represent  a good proof and reasoning that it is possible to contrast the evolution of bots also with state-of-art features and learning models.

\section*{Acknowledgements} Work partially supported by the Integrated Activity Project TOFFEe (TOols for Fighting FakEs) \url{https://toffee.imtlucca.it/}. 

\bibliographystyle{elsarticle-num}
\bibliography{biblio}
\newpage
\end{document}